# Classifying hydrogen-rich superconductors

E. F. Talantsev[1,2]

[1]M.N. Miheev Institute of Metal Physics, Ural Branch, Russian Academy of Sciences, 18, S. Kovalevskoy St., Ekaterinburg, 620108, Russia

[2]NANOTECH Centre, Ural Federal University, 19 Mira St., Ekaterinburg, 620002, Russia

E-mail: evgeny.talantsev@imp.uran.ru

*Abstract*

The era of near-room-temperature superconductivity started after experimental discovery by Drozdov *et al* (2015 *Nature* **525** 73) who found that compressed $H_3S$ exhibits superconducting transition at $T_c$ = 203 K. To date, the record near-room-temperature superconductivity stands with another hydrogen-rich highly compressed compound, $LaH_{10}$ (Somayazulu *et al* 2019 *Phys. Rev. Lett.* **122** 027001), which has critical temperature of $T_c >$ 240 K. In this paper, we analyse available upper critical field, $B_{c2}(T)$, data for $LaH_{10}$ (Drozdov *et al* 2019 *Nature* **569** 528) and report that this compound in all considered scenarios has the ratio of $T_c$ to the Fermi temperature, $T_F$, $0.009 < T_c/T_F < 0.038$, which is typical range for unconventional superconductors. In attempt to extend our finding, we examined experimental $B_{c2}(T)$ data for superconductors in the palladium-hydrogen ($PdH_x$) and thorium-hydrogen-deiterium (ThH-ThD) systems and surprisingly find that superconductors in both these systems also fall into unconventional superconductors band. Taking in account that $H_3S$ has the ratio of $0.012 < T_c/T_F < 0.039$ (Talantsev 2019 *Mod. Phys. Lett. B* **33** 1950195) we come to conclusion that in the Uemura plot all discovered to date hydrogen-rich superconductors, i.e. $Th_4H_{15}$-$Th_4D_{15}$, $PdH_x$, $H_3S$ and $LaH_{10}$ (in this list we do not include $NbTiH_x$, $PtH_x$, $SiH_4$, and $PH_3$ for which experimental data beyond $T_c$ are unknown), lie in same band as all unconventional superconductors, particularly heavy fermions, fullerenes, pnictides, and cuprates, and former should be classified as a new class of unconventional superconductors.

**I. Introduction**

The discovery of superconductivity in highly compressed H$_3$S with $T_c$ = 203 K by Drozdov *et al* [1] is the most fascinating breakthrough in superconductivity since epochal discovery of high-temperature superconductivity in cuprates by Bednoltz and Mueller [2]. Historical aspects of the problem can be found elsewhere [3-6].

Despite a fact that H$_3$S is widely classified as conventional [7] electron-phonon superconductor [1,3-6,8-13], we performed [14] analysis of experimental upper critical field data, $B_{c2}(T)$, reported by Mozaffari *et al* [15,16], and showed that H$_3$S is unconventional superconductor which lies in common unconventional superconductors trend band of the Uemura plot [17,18] together with other unconventional superconductors, i.e. heavy fermions, fullerenes, pnictides, and cuprates.

The concept of Uemura plot [18] is to represent all 32 classes of superconductors [19] by using two fundamental temperatures: one is the superconducting transition temperature, $T_c$ (usually, this is used as Y-axis), and another is the Fermi temperature, $T_F$ (usually, this is used as X-axis) [18]. In this representation all known unconventional superconductors fill a narrow band of $10^{-2} < T_c/T_F < 5 \cdot 10^{-2}$, while BCS electron-phonon mediated superconductors are located in the area of $T_c/T_F < 4 \cdot 10^{-4}$. We showed [14] that H$_3$S compressed at $P$ = 150 GPa and 155 GPa has $T_c/T_F$ ratio in the interval of $1.2 \cdot 10^{-2} < T_c/T_F < 3.9 \cdot 10^{-2}$, and, thus, based on general approach of Uemura *et al* [17,18], H$_3$S should be classified as unconventional superconductor.

Recently, further step towards room-temperature superconductivity was made by Somayazulu *et al* [20] who discovered near-room-temperature superconductivity in another highly-compressed hydrogen-rich compound of $LaH_{10}$. The latter exhibits transition temperature of $T_c \gtrsim 240\ K$ at external pressure in the range of $P$ = 150-200 GPa [20,21].

Drozdov *et al* [21] reported experimental upper critical field data for $LaH_{10}$ which we analyse in this paper and find that in all considered scenarios this compound has the ratio of superconducting energy gap, $\Delta(0)$, to the Fermi energy, $\varepsilon_F$, of $0.02 < \Delta(0)/\varepsilon_F < 0.07$, with respective ratio of $T_c$ to the Fermi temperature, $T_F$, $0.009 < T_c/T_F < 0.038$. As the result, in Uemura plot $LaH_{10}$ lies in the same band as all unconventional superconductors and falls just above another hydrogen-rich counterpart of $H_3S$ [14].

To prove that our findings in regard of $H_3S$ and $LaH_{10}$ are generic features of hydrogen-rich superconductors we re-examine the upper critical field data for the first discovered superhydride superconductor $Th_4H_{15}$ (by Satterthwaite and Toepke [22] in 1970), and, perhaps, the most experimentally studied to date hydrogen-rich superconductors in palladium-hydrogen-deuterium system, PdH-PhD, discovered by Skoskiewicz [23] and show that despite a fact that thorium-hydrogen-deuterium and palladium-hydrogen compounds have $T_c$ < 8.5 K, these low-temperature superconductors have similar ratios of $\Delta(0)/\varepsilon_F$ and $T_c/T_F$ as ones of its near-room-temperature superconducting counterparts.

We should note that the silane, $SiH_4$, was the first discovered by Eremets group highly-compressed hydrogen-rich superconductor with $T_c = 17\ K$ (observed at pressure of $P$ = 96-120 GPa) [24]. Covalent hydride phosphine, $PH_3$, is another hydrogen-rich superconductor in which superconductivity with $T_c \simeq 100\ K$ was discovered at $P \gtrsim 200\ GPa$ [25]. And recently, superconductivity in $PtH_x (x \cong 1)$ was discovered experimentally at $P$ = 30 GPa [26]. $NbTiH_x$ [27] is another hydrogen-rich superconductor which can be potentially considered. Unfortunately, for all of these compounds, fundamental experimental data beyond $T_c$ are unknown, and, thus, we were not able to analyse these materials in our consideration herein.

In result, we show that all discovered to date hydrogen-rich superconductors for which experimental data beyond $T_c$ are available, i.e. $Th_4H_{15}$-$Th_4D_{15}$, $PdH_x$, $H_3S$ and $LaH_{10}$, lie in the same band in the Uemura plot as all unconventional superconductors (particularly heavy fermions, fullerenes, pnictides, and cuprates) and thus, these superconductors should be classified as a new family of unconventional superconductors.

**II. The upper critical field models**

Ground state upper critical field, $B_{c2}(0)$, in the Ginzburg-Landau theory [28] is given by:

$$B_{c2}\left(\frac{T}{T_c}=0\right) = \frac{\phi_0}{2\cdot\pi\cdot\xi^2(0)}, \qquad (1)$$

where $\phi_0 = 2.068 \cdot 10^{-15}$ Wb is magnetic flux quantum, and $\xi(0)$ is the ground state coherence length. For the temperature dependent upper critical field data one of the most rigours model was proposed by Werthamer, Helfand, and Hohenberg (WHH) [29,30]:

$$ln\left(\frac{T}{T_c(B=0)}\right) = \psi\left(\frac{1}{2}\right) - \psi\left(\frac{1}{2} + \frac{\hbar \cdot D \cdot B_{c2}(T)}{2\cdot\phi_0 \cdot k_B \cdot T}\right) \qquad (2)$$

where $\psi$ is digamma function, $D$ is the diffusion constant of the normal conducting electrons/holes. However, application of this model for the analysis of experimental $B_{c2}(T)$ data requires very often and uniformly measured dataset which covers the whole temperature range of $0 < T < T_c$ with approximate step of $(0.02 - 0.05) \cdot T_c$ between data points. In real world experiments, this condition is usually impossible to achieve due to either experimental limitation to cool sample down to low enough reduced temperatures, $T/T_c$, either to create reasonably high applied magnetic field, $B_{appl}/B_{c2}(0)$, or, in most cases, the both.

Based on this, the analysis of $B_{c2}(T)$ data, as a rule, performs by the utilization of other approaches, different from WHH [29,30], which were developed for real world experiments, i.e. when $B_{c2}(T)$ dataset is limited by measurements performed at high reduced temperatures, $\frac{1}{2} \lesssim \frac{T}{T_c}$. We stress that this is entire case for highly compressed $LaH_{10}$ for which $B_{c2}(T)$ data is available to date only in narrow temperature range of $0.96 \leq \frac{T}{T_c} \leq 1.0$ [21]. Primary reason for this is that available in experiment applied magnetic field was limited by $B_{appl} \leq 9\,T$ [21], which is typical for modern labs (for instance, conventional PPMS systems provide this). However, if even world top quasi-DC magnetic field facility will be in use [15,16,31] and maximum applied magnetic field of $B_{appl} = 62 - 65\,T$ will be utilized, then based on estimated value for $B_{c2}(0)$ = 90-140 T (given by Drozdov *et al* [21]) for LaH$_{10}$, it is unlikely that $B_{c2}(T)$ measurements will be possible to perform at reduced temperature lower than $0.7 \leq \frac{T}{T_c}$, i.e., $170\,K \leq T$.

The reality is that robust use of WHH model [29,30] for highly compressed LaH$_{10}$ (and, probably, for others homological compounds of La$_n$H$_m$ series [32,33]) will be unlikely in near decades. Thus, different models need to be used to analyse experimental $B_{c2}(T)$ data for near-room-temperature superconductors. One of the approach, which was actually proposed by Werthamer, Helfand, and Hohenberg [29,30], is extrapolative expression:

$$B_{c2}(0) = \frac{\phi_0}{2\cdot\pi\cdot\xi^2(0)} = -0.693 \cdot T_c \cdot \left(\frac{dB_{c2}(T)}{dT}\right)_{T\sim T_c} \qquad (3)$$

In this paper, we will designate Eq. 3 as WHH model.

Another model, which is also based on Werthamer, Helfand, and Hohenberg primary idea [29,30], but one accurately generates full $B_{c2}(T)$ extrapolative curve from experimental data measured at high reduced temperatures, $T/T_c$, was developed by Baumgartner *et al* [34]:

$$B_{c2}(T) = B_{c2}(0) \cdot \left(\frac{\left(1-\frac{T}{T_c}\right)-0.153\cdot\left(1-\frac{T}{T_c}\right)^2-0.152\cdot\left(1-\frac{T}{T_c}\right)^4}{0.693}\right) =$$

$$= \frac{\phi_0}{2\cdot\pi\cdot\xi^2(0)} \cdot \left(\frac{\left(1-\frac{T}{T_c}\right)-0.153\cdot\left(1-\frac{T}{T_c}\right)^2-0.152\cdot\left(1-\frac{T}{T_c}\right)^4}{0.693}\right) \qquad (4)$$

We will designate this model as B-WHH model.

In addition, we will use classical two-fluid Gorter-Casimir model (GC model) [35,36]:

$$B_{c2}(T) = B_{c2}(0) \cdot \left(1 - \left(\frac{T}{T_c}\right)^2\right) = \frac{\phi_0}{2 \cdot \pi \cdot \xi^2(0)} \cdot \left(1 - \left(\frac{T}{T_c}\right)^2\right) \qquad (5)$$

which is in a wide use too [1,15,21,37,38]. We note, that Drozdov *et al* [1], Mozaffari *et al* [15,16] and Drozdov *et al* [21] designate Eq. 5 as the Ginzburg-Landau theory equation [28]. This is incorrect, because the latter is [28,35]:

$$B_{c2}(T) = B_{c2}(0) \cdot \left(1 - \left(\frac{T}{T_c}\right)\right) = \frac{\phi_0}{2 \cdot \pi \cdot \xi^2(0)} \cdot \left(1 - \left(\frac{T}{T_c}\right)\right) \qquad (6)$$

There is also $B_{c2}(T)$ model proposed by Gor'kov [39] which was written in analytical form by Jones *et al.* [40]:

$$B_{c2}(T) = B_{c2}(0) \cdot \left(\frac{1.77 - 0.43 \cdot \left(\frac{T}{T_c}\right)^2 + 0.07 \cdot \left(\frac{T}{T_c}\right)^4}{1.77}\right) \cdot \left[1 - \left(\frac{T}{T_c}\right)^2\right] =$$

$$\frac{\phi_0}{2 \cdot \pi \cdot \xi^2(0)} \cdot \left(\frac{1.77 - 0.43 \cdot \left(\frac{T}{T_c}\right)^2 + 0.07 \cdot \left(\frac{T}{T_c}\right)^4}{1.77}\right) \cdot \left[1 - \left(\frac{T}{T_c}\right)^2\right] \qquad (7)$$

### III. $Th_4H_{15}$-$Th_4D_{15}$ superconductors in Uemura plot

We start our consideration with the first discovered superhydride superconductors, i.e., $Th_4H_{15}$ and $Th_4D_{15}$ [22]. We should note, that already in the first paper of 1970, Satterthwaite and Toepke [22] reported the absence of the isotope effect in ThH-ThD system:

$$T_c \cdot M^\alpha = const. \qquad (8)$$

where $M$ is isotope mass, and $\alpha \approx 1/2$ for weak-coupling limit of BCS theory [7], which is one of indispensable fundamental feature of electron-phonon mediated superconductivity [7]. Later, Stritzker and Buckel [41] experimentally found that the isotope effect in the palladium-hydrogen-deuterium (PdH-PdD) system has opposite sign (so called, reverse isotope effect). Yussouff *et al.* [42] extended this discovery on palladium-hydrogen-deuterium-tritium system (PdH-PdD-PdT). This reverse isotope effect in PdH-PdD-PdT system is still under wide discussion [43,44]. In regard of ThH-ThD system, detailed studied by Caton and Satterthwaite [45] showed that superconductors in thorium-hydrogen-deuterium (ThH-ThD) system have also reverse isotope effect.

From the author knowledge, available to date experimental data for the upper critical field, $B_{c2}(T)$, are limited by values reported by Satterthwaite and Toepke [22]. The authors reported for both, $Th_4H_{15}$ and $Th_4D_{15}$, compounds:

$$B_{c2}(T\sim 0) = 2.5 - 3.0\ T. \tag{9}$$

From these values, the ground state coherence length, $\xi(0)$, for $Th_4H_{15}$ and $Th_4D_{15}$ phases, can be derived as following:

$$\xi(0) = 11.0 \pm 0.5\ nm \tag{10}$$

Miller *et al.* [46] for both phases reported the BCS ratio within a range:

$$\alpha = \frac{2 \cdot \Delta(0)}{k_B \cdot T_c} = 3.42 - 3.47. \tag{11}$$

By utilizing superconducting transition temperature for $Th_4H_{15}$ and $Th_4D_{15}$ phases [22]:

$$T_c = 8.20 \pm 0.15\ K \tag{12}$$

one can deduce ground state superconducting energy gap:

$$\Delta(0) = 1.22 \pm 0.03\ meV \tag{13}$$

and by using well-known BCS expression [7]:

$$\xi(0) = \frac{\hbar \cdot v_F}{\pi \cdot \Delta(0)} \tag{14}$$

where $\hbar = h/2\pi$ is reduced Planck constant, one can calculate the Fermi velocity, $v_F$, in $Th_4H_{15}$ and $Th_4D_{15}$ phases:

$$v_F = \pi \cdot \frac{\xi(0) \cdot \Delta(0)}{\hbar} = (6.4 \pm 0.2) \cdot 10^4\ m/s \tag{15}$$

To place $Th_4H_{15}$ and $Th_4D_{15}$ phases in the Uemura plot [18], we need to make assumption about the effective charge carrier mass, $m^*_{eff}$, to calculate the Fermi temperature, $T_F$:

$$T_F = \frac{\varepsilon_F}{k_B} = \frac{m^*_{eff} \cdot v_F^2}{2 \cdot k_B} \tag{16}$$

Due to there is no any available experimental values to date for $Th_4H_{15}$ and $Th_4D_{15}$ phases we used the simplest assumption:

$$m^*_{eff} = 1.0 \cdot m_e \qquad (17)$$

and one can calculate the Fermi temperature, $T_F$:

$$T_F = \frac{\varepsilon_F}{k_B} = \frac{m^*_{eff} \cdot v_F^2}{2 \cdot k_B} = 137 \pm 9\ K \qquad (18)$$

and the ratio:

$$\frac{T_c}{T_F} = 0.060 \pm 0.005 \qquad (19)$$

which places both $Th_4H_{15}$ and $Th_4D_{15}$ phases in the Uemura plot in the upper boarder of the unconventional superconductor (Fig. 1).

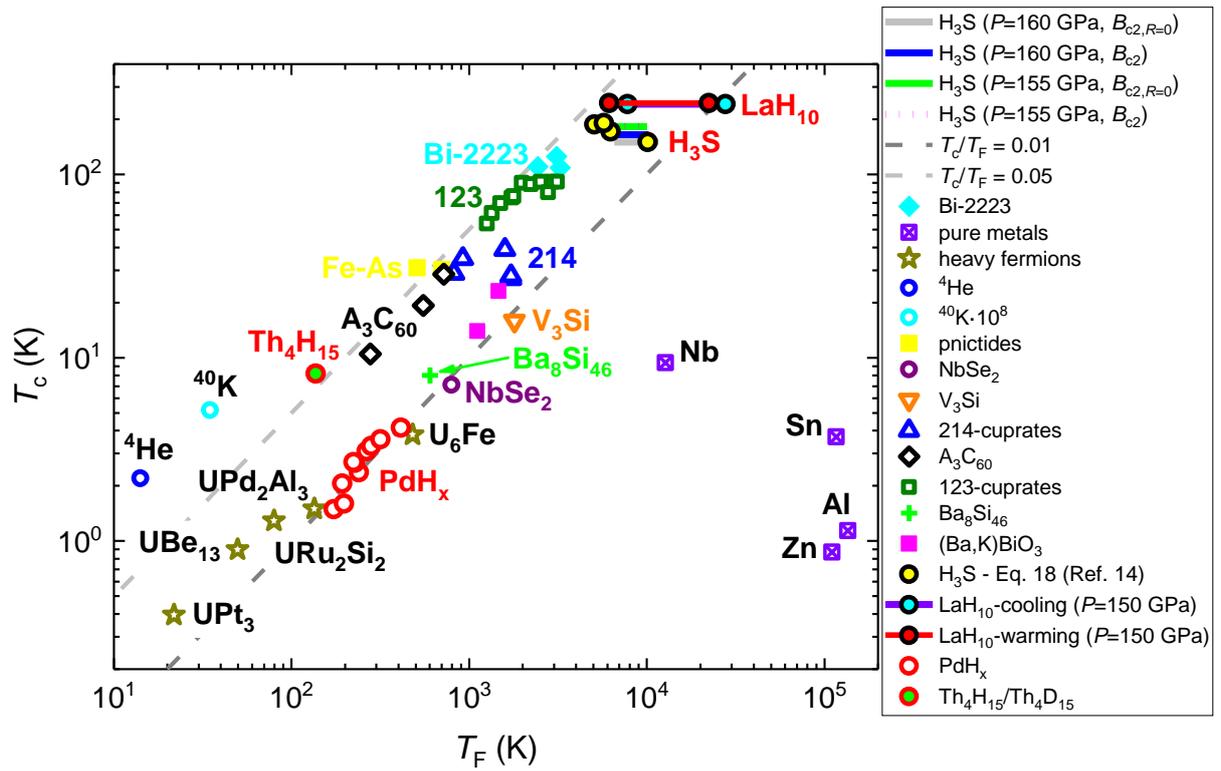

**Figure 1.** A plot of $T_c$ versus $T_F$ obtained for most representative superconducting families including $PdH_x$, $H_3S$ and $LaH_{10}$. Data was taken from Uemura [18], Ye *et al.* [47], Qian *et al.* [48], Hashimoto *et al.* [49] and Ref. 14.

**IV. PdH$_x$ superconductors in Uemura plot**

Surprisingly enough, to author's knowledge, as thorium-hydrogen-deuterium superconductors discovered by Satterthwaite and Toepke [22] in 1970, as palladium-hydrogen-deuterium superconductors discovered by Skoskiewicz [23] in 1973 have never been located in the Uemura plot [18]. To do this, we utilize results of systematic studies of the upper critical field for palladium-hydrogen superconductors reported by Balbaa and Manchester [50]. We show their results (i.e., $T_c$ and $B_{c2}(0)$) in Table I together with calculated by using Eq. 1 values for $\xi(0)$.

Skoskiewicz [23] reported that PdH$_x$ superconductors have the ratio:

$$\alpha = \frac{2 \cdot \Delta(0)}{k_B \cdot T_c} = 3.7 \qquad (20)$$

where $\Delta(0)$ is ground state superconducting energy gap, $k_B$ is the Boltzmann constant, which is very close to BCS weak-coupling limit of 3.53 [7]. Sansores *et al.* [51] performed the analysis of experimental electronic specific heat jump at $T_c$, $\Delta C/C$, and confirmed the weak-coupling pairing strength in PdH$_x$ and PdD$_x$ by reporting:

$$\frac{\Delta C}{C} = 1.44 - 1.59 \qquad (21)$$

which is very close to BCS weak-coupling limit of 1.43 [7]. Thus, we use $\alpha$ value reported by Skoskiewicz (Eq. 8) [23] in our calculations below.

Bambakidis *et al.* [52] reported that PdD$_x$ compounds have charge carrier effective mass of:

$$m^*_{eff} = 0.49 \cdot m_e \qquad (22)$$

where $m_e$ is electron mass, which we use in calculations below for PdH$_x$ superconductors.

By using Eqs. 9-19 we calculate value which presented in Table I and Fig. 1.

**Table I.** Deduced parameters for PdH$_x$ superconductors based on data reported by Balbaa and Manchester [50]. We assumed that $\frac{2\cdot\Delta(0)}{k_B\cdot T_c} = 3.7$ [23] and electron effective mass is $m^*_{eff} = 0.49 \cdot m_e$ [52]. Maximal and minimal $T_c/T_F$ values are in red bold.

| $x$ = H/Pd | $T_c$ (K) | $B_{c2}(0)$ (mT) | Deduced $\xi(0)$ (nm) | $v_F$ ($10^5$ m/s) | $\Delta(0)$ meV | $\varepsilon_F$ meV | $\Delta(0)/\varepsilon_F$ | $T_F$ | $T_c/T_F$ |
|---|---|---|---|---|---|---|---|---|---|
| 0.821 | 1.488 | 39.5 | 91.3 | 1.03 | 0.237 | 14.9 | 0.016 | 173 | 0.009 |
| 0.826 | 1.600 | 40.0 | 90.8 | 1.11 | 0.255 | 15.0 | 0.015 | 198 | **0.008** |
| 0.843 | 2.061 | 68.0 | 69.6 | 1.09 | 0.329 | 16.6 | 0.020 | 193 | 0.011 |
| 0.852 | 2.365 | 72.0 | 67.6 | 1.22 | 0.377 | 20.6 | 0.018 | 239 | 0.010 |
| 0.862 | 2.672 | 96.5 | 58.4 | 1.19 | 0.426 | 19.6 | 0.022 | 228 | **0.012** |
| 0.863 | 2.695 | 100.0 | 57.4 | 1.18 | 0.430 | 19.3 | 0.022 | 224 | **0.012** |
| 0.875 | 3.090 | 111.0 | 54.5 | 1.28 | 0.493 | 22.9 | 0.022 | 265 | **0.012** |
| 0.881 | 3.305 | 120.0 | 52.4 | 1.32 | 0.527 | 24.2 | 0.022 | 281 | **0.012** |
| 0.887 | 3.590 | 125.0 | 51.3 | 1.40 | 0.572 | 27.4 | 0.021 | 317 | 0.011 |
| 0.905 | 4.158 | 129.0 | 50.5 | 1.60 | 0.663 | 35.6 | 0.019 | 413 | 0.010 |

It can be seen (Fig. 1) that PdH$_x$ falls just next to heavy fermion superconductors and thus these hydrogen-rich superconductors should be classified as unconventional superconductors.

### V. Deduced $B_{c2}(0)$ and $\xi(0)$ for compressed LaH$_{10}$

In Fig. 2,a we show raw $B_{c2}(T)$ data for compressed LaH$_{10}$ measured at the "cooling" stage (see for details Ref. 21). $B_{c2}(T)$ dataset was deduced from $R(T)$ curves showed in Fig. 2(a) of Drozdov *et al* [21] by using 50% of the normal state resistance criterion (raw data are given in Supplementary Table I). We note that Drozdov *et al* [21] also used this criterion to deduce $B_{c2}(T)$ and they also fitted data to Eq. 5 in their Fig. 2(b). In our Fig. 2,a we show fits to Eqs. 3-5,7 with deduced values collected in Table II.

For so-called "warming" stage, Drozdov *et al* [21] registered different $R(T)$ curves (Fig. 2(a) [21]) which we process in the same way and fit to Eqs. 3-5,7 (Fig. 2,b) (raw data are given in Supplementary Table II). Deduced $B_{c2}(0)$ and $\xi(0)$ values are collected in Table II.

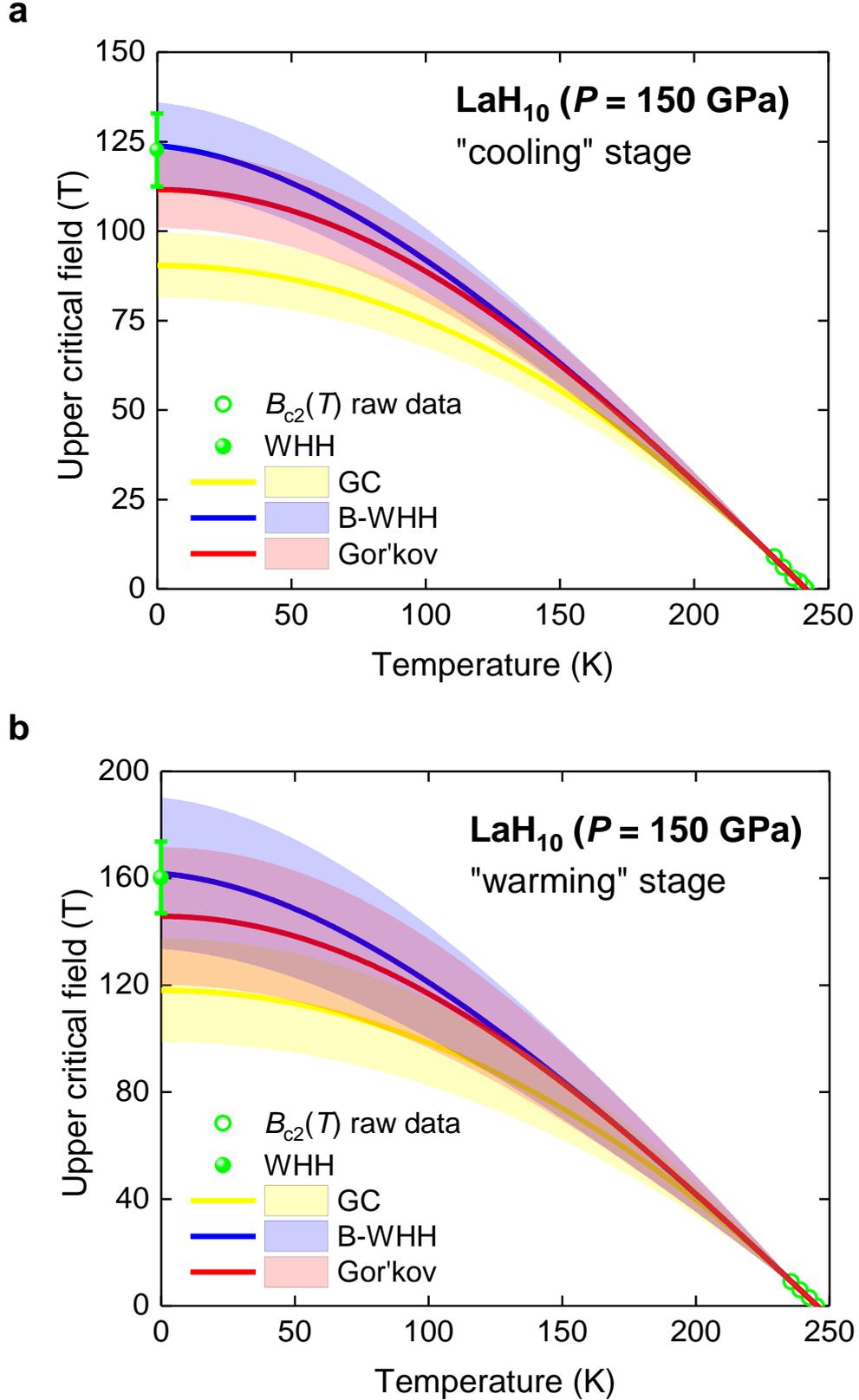

**Figure 2.** Superconducting upper critical field, $B_{c2}(T)$, data (magenta circles) and fits to four different model (Eqs. 3-5,7) for $LaH_{10}$ compressed at pressure $P = 150$ GPa (raw data are from Ref. 21) at (a) "cooling" and (b) "warming" stages. 95% confidence bars are shown.

## IV. Compressed LaH₁₀ in Uemura plot

Kruglov *et al.* [33] performed first principles calculations and came to conclusion that compressed LaH₁₀ is strong coupled superconductor with the ratio:

$$\alpha = \frac{2 \cdot \Delta(0)}{k_B \cdot T_c} = 5.00 - 5.55 \quad (23)$$

where $\Delta(0)$ is ground state superconducting energy gap, $k_B$ is the Boltzmann constant. Despite a fact that the upper boundary for α is the highest ever reported for electron-phonon mediated superconductors [9] we use this value as maximal α amplitude in our calculations below (Table II and Fig. 1).

We note that first principles calculations performed for H₃S also showed high values for the ratio (references can be found elsewhere [4,9]):

$$\alpha = \frac{2 \cdot \Delta(0)}{k_B \cdot T_c} = 4.5 - 4.7 \quad (24)$$

**Table II.** Deduced parameters for LaH₁₀ superconductor subjected to external pressure of $P = 150$ GPa. We assumed that electron effective mass is $m^*_{eff} = 3.0 \cdot m_e$ [58]. Maximal and minimal $T_c/T_F$ values are in red bold.

| Stage | Model | Deduced $T_c$ (K) | Deduced $\xi(0)$ (nm) | Assumed $\frac{2 \cdot \Delta(0)}{k_B \cdot T_c}$ | $v_F$ (10⁵ m/s) | $\Delta(0)$ meV | $\varepsilon_F$ eV | $\Delta(0)/\varepsilon_F$ | $T_F$ (10³ K) | $T_c/T_F$ |
|---|---|---|---|---|---|---|---|---|---|---|
| *cooling* | B-WHH | 241.7 ± 0.2 | 1.63 ± 0.03 | 3.53 | 2.9 ± 0.1 | 36.8 ± 0.1 | 0.70 ± 0.03 | 0.053 ± 0.002 | 8.1 ± 0.3 | 0.030 ± 0.001 |
| | | | | 5.55 | 4.5 ± 0.1 | 57.8 ± 0.1 | 1.72 ± 0.06 | 0.034 ± 0.001 | 20.0 ± 0.8 | 0.012 ± 0.001 |
| | GC | 241.6 ± 0.2 | 1.91 ± 0.04 | 3.53 | 3.4 ± 0.1 | 36.7 ± 0.1 | 0.96 ± 0.04 | 0.038 ± 0.002 | 11.1 ± 0.5 | 0.022 ± 0.001 |
| | | | | 5.55 | 5.3 ± 0.1 | 57.8 ± 0.1 | 2.40 ± 0.10 | 0.024 ± 0.002 | 27.5 ± 0.8 | **0.009 ± 0.001** |
| | G | 241.7 ± 0.2 | 1.72 ± 0.04 | 3.53 | 3.0 ± 0.1 | 36.8 ± 0.1 | 0.78 ± 0.04 | 0.047 ± 0.003 | 9.0 ± 0.4 | 0.027 ± 0.001 |
| | | | | 5.55 | 4.7 ± 0.1 | 57.8 ± 0.1 | 1.92 ± 0.09 | 0.030 ± 0.002 | 22.3 ± 1.1 | 0.011 ± 0.001 |
| *warming* | B-WHH | 245.3 ± 0.2 | 1.43 ± 0.03 | 3.53 | 2.6 ± 0.1 | 37.3 ± 0.1 | 0.55 ± 0.02 | 0.068 ± 0.003 | 6.4 ± 0.3 | **0.038 ± 0.002** |
| | | | | 5.55 | 4.0 ± 0.1 | 58.7 ± 0.1 | 1.37 ± 0.06 | 0.043 ± 0.002 | 15.9 ± 0.7 | 0.016 ± 0.001 |
| | GC | 245.3 ± 0.2 | 1.67 ± 0.03 | 3.53 | 3.0 ± 0.1 | 37.3 ± 0.1 | 0.75 ± 0.03 | 0.050 ± 0.003 | 8.8 ± 0.4 | 0.028 ± 0.001 |
| | | | | 5.55 | 4.7 ± 0.1 | 58.7 ± 0.1 | 1.86 ± 0.09 | 0.032 ± 0.002 | 21.6 ± 0.9 | 0.011 ± 0.001 |
| | G | 245.3 ± 0.3 | 1.50 ± 0.03 | 3.53 | 2.7 ± 0.1 | 37.3 ± 0.1 | 0.61 ± 0.03 | 0.061 ± 0.004 | 7.1 ± 0.4 | 0.035 ± 0.002 |
| | | | | 5.55 | 4.2 ± 0.1 | 58.7 ± 0.1 | 1.50 ± 0.06 | 0.039 ± 0.003 | 17.5 ± 0.7 | 0.014 ± 0.001 |

From other hand, Hirsch and Marsiglio [53], Souza and Marsiglio [54], Harshman and Fiory [55], Kaplan and Imry [56] proposed different models for the superconductivity in H$_3$S either within classical BCS approach [56], either based on new concepts, and, for instance, Kaplan and Imry [56] showed that their model gives α within weak-coupling BCS limit:

$$\alpha = \frac{2 \cdot \Delta(0)}{k_B \cdot T_c} = 3.53 \quad (25)$$

This α value is in a good agreement with ones we deduced from experimental $B_{c2}(T)$ [14] and the self-field critical current density, $J_c(sf,T)$, data [57]. Assuming that both hydrogen-rich counterparts (H$_3$S and LaH$_{10}$) have the same origin for near-room-temperature superconductivity, we calculated parameters for the latter (Table II and Fig. 1) by using weak-coupling limit of BCS (Eq. 25) as the lowest value for α.

Kostrzewa *et al.* [58] reported that at $P = 150$ GPa compressed LaH$_{10}$ has charge carrier effective mass of:

$$m^*_{eff} = (2.9 - 3.2) \cdot m_e \approx 3.0 \cdot m_e \quad (26)$$

where $m_e$ is electron mass. This value is very close to $m^*_{eff} = 2.76 \cdot m_e$ calculated for compressed H$_3$S by Durajski [11]. For the simplicity, in our calculations below, for compressed LaH$_{10}$ we use rounded value of $m^*_{eff} = 3.0 \cdot m_e$.

We perform the same calculation routine to one as we did for PdH$_x$ and calculated values are given in Table II and displayed in Fig. 1. Deduced $T_c/T_F$ values are in the range of $0.9 \cdot 10^{-2} < T_c/T_F < 4.0 \cdot 10^{-2}$. It can be seen (Fig. 1) that LaH$_{10}$ falls just above another hydrogen-rich counterpart, H$_3$S, and both near-room-temperature superconductors are placed in unconventional superconductors band together with heavy fermions, PdH$_x$, fullerenes, pnictides and cuprates.

More evidently our primary conclusion that H₃S and LaH₁₀ are unconventional superconductors can be understood if one substitutes Eqs. 9-19 in the equation Eq. 1 and obtains:

$$B_{c2}\left(\frac{T}{T_c} = 0\right) = \frac{\pi \cdot \phi_0 \cdot k_B}{16 \cdot \hbar^2} \cdot m^*_{eff} \cdot \alpha^2 \cdot \left(\frac{T_c}{T_F}\right) \cdot T_c, \tag{27}$$

If LaH₁₀ will be a conventional superconductor with $T_c/T_F = 8.4 \cdot 10^{-6}$ (this is the ratio of electron-phonon mediated Al [18]), then in accordance with Eq. 27:

$$B_{c2}\left(\frac{T}{T_c} = 0\right) = 70 \; mT \tag{28}$$

If $T_c/T_F = 7.4 \cdot 10^{-4}$ (which is the ratio of electron-phonon mediated Nb [18]), then:

$$B_{c2}\left(\frac{T}{T_c} = 0\right) = 6.2 \; T \tag{29}$$

We note, that experimental $B_{c2}(T)$ value for LaH₁₀ [21] is:

$$B_{c2}\left(\frac{T}{T_c} = 0.96\right) = 9.0 \; T \tag{30}$$

Thus, Eqs. 27-30 show that despite a fact that currently $B_{c2}(T)$ dataset for LaH₁₀ is available only at temperatures near $T_c$ (i.e., $0.96 \leq \frac{T}{T_c} \leq 1.0$), this dataset is already enough to make a conclusion that LaH₁₀ is unconventional superconductor.

Similar calculations for H₃S (based on experimental [15,16] and deduced [14] values) are:

$$B_{c2}\left(\frac{T}{T_c} = 0\right) < 45 \; mT \quad (\text{H}_3\text{S for } T_c/T_F = 8.4 \cdot 10^{-6}) \tag{31}$$

$$B_{c2}\left(\frac{T}{T_c} = 0\right) < 3.9 \; T \quad (\text{H}_3\text{S for } T_c/T_F = 7.4 \cdot 10^{-4}). \tag{32}$$

These values (if even we will make comparison of ones with very strictly defined $B_{c2}(T)$ (at $R = 0 \; \Omega$) and pressure of $P = 160$ GPa (which is shifted from optimal pressure of $P = 155$ GPa)) are at least in one order of magnitude lower than experimental data:

$$B_{c2}\left(\frac{T}{T_c} \approx \frac{1}{3}\right) = 65 \; T \quad (\text{H}_3\text{S at } P = 160 \text{ GPa}). \tag{33}$$

Eqs. 27-33 show that both near-room-temperature superconductors, $H_3S$ and $LaH_{10}$, cannot be classified as conventional superconductors, despite a fact that astonishing Ashcroft's prediction [59,60] about near-room-temperature superconductivity in hydrogen-rich compounds was based on BCS theory.

In overall, all extrapolated $B_{c2}(0)$ values for $LaH_{10}$ (Fig. 2) are well below the Pauli depairing field of:

$$B_p(0) = \frac{2 \cdot \Delta(0)}{g \cdot \mu_B} = \alpha \cdot \frac{k_B \cdot T_c}{g \cdot \mu_B} = 630 - 1{,}000\ T \gg B_{c2}(0) \tag{34}$$

where $g = 2$ and $\mu_B = \frac{e \cdot \hbar}{2 \cdot m_e}$ is the Bohr magneton. According to Gor'kov's note [61], Eq. 34 means that the mean-free path, $l$, of the electrons is large compared with the coherence length, $\xi(0)$:

$$l \gg \xi(T) > \xi(0) \sim 1.4 - 1.9\ nm \tag{35}$$

However, the lower limit for the mean-free path, $l$, is only in a few times exceeded the lattice constants for any structural unit cell of $LaH_{10}$ either proposed by first principles calculations [33], either deduced by the fits of experimental data to *Fm-3m* symmetry ($a = 0.51019(5)$ nm [21]) or *P4/nmm* symmetry ($a = 0.37258(6)$ nm, c = $0.50953(12)$ nm [21]). This means that charge carrier scattering is large in $LaH_{10}$.

### V. Thermodynamic fluctuations in compressed $LaH_{10}$

In our previous papers [57,62] we answered a question about possible limitation of $T_c$ in compressed $H_3S$ by the thermodynamic fluctuations of the order parameter. There are phase [63] and amplitude [64] fluctuations of the order parameter in superconductors. These fluctuations for the case of three dimensional (3D) superconductors (like, $Th_4H_{15}$, $PdH_x$, $H_3S$ and $LaH_{10}$) have characteristic temperatures:

$$T_{fluc,phase} = \frac{0.55 \cdot \phi_0^2}{\pi^{3/2} \cdot \mu_0 \cdot k_B} \cdot \frac{1}{\kappa^2 \cdot \xi(0)} \tag{36}$$

$$T_{fluc,amp} = \frac{\phi_0^2}{12 \cdot \pi \cdot \mu_0 \cdot k_B} \cdot \frac{1}{\kappa^2 \cdot \xi(0)} \qquad (37)$$

where $\kappa = \lambda(0)/\xi(0)$ is the Ginzburg-Landau parameter, and $\lambda(0)$ is the ground state London penetration depth.

Due to for $Th_4H_{15}$ and $Th_4D_{15}$ phases the Ginzburg-Landau parameter is unknown we do not consider this interesting case herein.

For $PdH_x$ system, $\kappa$ was found to be $\kappa = 1.2$-$2.0$ [50], and calculated values for $T_{fluc,phase}$ and $T_{fluc,amp}$ show that thermodynamic fluctuations for superconductors in palladium-hydrogen system are negligible low (Table III).

In our recent paper [14], we deduced $\xi(0)$ in $H_3S$ more accurately in comparison with values we used to calculate $T_{fluc,phase}$ and $T_{fluc,amp}$ in our previous papers [57,62]. By taking in account that $H_3S$ exhibits $\kappa = 88$-$105$ [57,62,65], we revisited $T_{fluc,phase}$ and $T_{fluc,amp}$ at two applied pressures of $P = 155$ GPa and $160$ GPa (Table III). At each pressure, we used maximal and minimal deduced $\xi(0)$ values to show the range of variation for $T_{fluc,phase}$ and $T_{fluc,amp}$.

To calculate $T_{fluc,phase}$ and $T_{fluc,amp}$ (Eqs. 36,37) for compressed $LaH_{10}$ there is a need to make an assumption about the value for Ginzburg-Landau parameter, $\kappa$. Our approach is based on an assumption that $\kappa$ for $LaH_{10}$ will be not much different from $\kappa$ for $H_3S$ and other unconventional superconductors (primarily, pnictides and cuprates) which have values within a range of $\kappa = 60$-$120$ [65-71]. Calculated $T_{fluc,phase}$ and $T_{fluc,amp}$ for $LaH_{10}$ are given in Table III.

Examination of the values in Table III leaded us to an important finding that both near-room-temperature superconductors have very large fluctuations of the order parameter amplitude, which, in some scenarios, have characteristic temperatures, $T_{fluc,amp}$, of only about 15% above observed transition temperature, $T_c$. This is similar to cuprates and pnictides which have large fluctuations of order parameter phase [57,62] (which causes the suppression of the observed $T_c$ by about 30% from its mean-field value, $T_c^{mean-field}$ [57,62]).

**Table III.** Calculated fluctuation temperatures for H$_3$S (at $P$ = 155 and 160 GPa) and LaH$_{10}$ ($P$ = 150 GPa). Assumed electron effective masses are $m_{eff}^* = 2.76 \cdot m_e$ [11] for H$_3$S and $m_{eff}^* = 3.0 \cdot m_e$ for LaH$_{10}$ [50]. The largest ratios for $\frac{T_c}{T_{fluc,phase}}$ and $\frac{T_c}{T_{fluc,amp}}$ are marked in red bold. Deduced $T_c$ and ξ(0) values for H$_3$S are from Ref. 14.

| Material | Stage | Pressure (GPa) | Deduced $T_c$ (K) | Deduced ξ(0) (nm) | Assumed κ | $T_{fluc,phase}$ (K) | $T_{fluc,amp}$ (K) | $T_c/T_{fluc,phase}$ | $T_c/T_{fluc,amp}$ |
|---|---|---|---|---|---|---|---|---|---|
| PdH$_x$ | N/A | N/A | 4.16 | 50.5 | 2 | 1.2·10$^5$ | 3.2·10$^4$ | 3.5·10$^{-5}$ | 1.3·10$^{-4}$ |
| H$_3$S | N/A | 155 | 182 ± 1 | 1.97 ± 0.02 | 88 | 1590 ± 20 | 428 ± 5 | 0.114 ± 0.001 | 0.43 ± 0.01 |
| | | | | | 105 | 1120 ± 11 | 304 ± 5 | **0.163 ± 0.002** | **0.61 ± 0.01** |
| | | | 189 ± 1 | 1.68 ± 0.01 | 88 | 1870 ± 10 | 502 ± 3 | 0.101 ± 0.001 | 0.38 ± 0.01 |
| | | | | | 105 | 1313 ± 11 | 353 ± 5 | 0.144 ± 0.001 | 0.54 ± 0.01 |
| | | 160 | 150 ± 3 | 2.67 ± 0.05 | 88 | 1176 ± 23 | 316 ± 6 | 0.128 ± 0.003 | 0.48 ± 0.01 |
| | | | | | 105 | 826 ± 16 | 222 ± 4 | **0.182 ± 0.004** | **0.68 ± 0.02** |
| | | | 172 ± 2 | 2.06 ± 0.01 | 88 | 1524 ± 8 | 410 ± 2 | 0.113 ± 0.001 | 0.42 ± 0.01 |
| | | | | | 105 | 1071 ± 5 | 288 ± 2 | 0.161 ± 0.002 | 0.60 ± 0.01 |
| LaH$_{10}$ | *cooling* | 150 | 241.7 ± 0.2 | 1.63 ± 0.03 | 60 | 4150 ± 70 | 1113 ± 30 | 0.058 ± 0.001 | 0.21 ± 0.01 |
| | | | | | 120 | 1036 ± 20 | 278 ± 6 | **0.233 ± 0.004** | **0.87 ± 0.02** |
| | *warming* | 150 | 245.3 ± 0.2 | 1.43 ± 0.03 | 60 | 4720 ± 100 | 1270 ± 30 | 0.052 ± 0.001 | 0.19 ± 0.01 |
| | | | | | 120 | 1180 ± 25 | 317 ± 7 | 0.208 ± 0.005 | 0.77 ± 0.02 |

## VI. Self-field critical currents in compressed LaH$_{10}$

We need to stress, that there is fundamental limit to obtain answers on many important questions in regards of H$_3$S and LaH$_{10}$ by performing the upper critical field studies, because experimentally available magnetic fields, $B_{appl}$, even, if top world facilities will be in use [15,16,31], are too low for these materials. Thus, there is a need to find different experimental techniques to reveal the nature of superconductivity in these and, perhaps, many others near-room-temperature hydrogen-based superconductors. One possible way to perform this is to study temperature dependent self-field critical currents, $I_c(sf,T)$ [37,65], from which several fundamental parameters of the superconductor, i.e., the ground state superconducting energy gap, $\Delta(0)$, the ground state London penetration depth, $\lambda(0)$, and relative jump in the specific heat at $T_c$, $\Delta C/C$, can be deduced. We already showed that this approach works for compressed $H_3S$ [57] by performing analysis of the self-field magnetization critical current densities reported by Drozdov *et al.* [1].

We note that experimental technique to perform critical current measurements in ultrahigh-pressure diamond cells is under developing for about twenty years [72-75] and fundamental possibility to measure $I_c(sf,T)$ in $LaH_{10}$ has been already demonstrated by Somayazulu *et al* [20] in their Fig. 4 and Supplementary Information. However, $I_c(sf,T)$ measurement techniques (and particularly inside of ultrahigh-pressure diamond cells) need to be further developed, because these measurements require great precaution due to the danger of sample "burning" during transport current pulse [76-78], and $I_c(sf,T)$ data collecting is still state-of-art [79,80]. For instance, $LaH_{10}$ samples degradation, after $I_c(sf,T)$ measurements reported by Somayazulu *et al* [20], is more likely originated from the sample "burning" under transport current flow. In this regard, recently registered effect [81] of the linearization of surface magnetic field, $B_{surf}(I)$, vs transport current rise at the onset of power dissipation can be considered as a new option. We showed recently, that this effect [81,82] works when external magnetic field, $B_{appl}$, is applied to the sample [83,84].

There are also two powerful optical spectroscopy techniques, one is the Raman spectroscopy (which is under on-going developing to be used in ultrahigh-pressure diamond cells for last decades [85-88]), and another is the angle resolved photoemission spectroscopy (ARPES) which was recently advanced [89] to be applicable to study superconducting films under transport current flow.

## VII. Conclusions

In this paper we analyse experimental $B_{c2}(T)$ data of near-room-temperature superconductor, $LaH_{10}$ (recently reported by Drozdov *et al.* [21]), of palladium-hydrogen superconductors, $PdH_x$ (reported by Balbaa and Manchester [50]) and of $Th_4H_{15}$-$Th_4D_{15}$ (reported by Satterthwaite and Toepke [22]). We come to conclusion that all discovered to date hydrogen-rich superconductors for which fundamental superconducting parameters beyond $T_c$ were measured (in this list we do not include $NbTiH_x$, $PtH_x$, $SiH_4$ and $PH_3$ for which only experimental $T_c$ vs pressure are known), i.e., $Th_4H_{15}$-$Th_4D_{15}$, $PdH_x$, $H_3S$ and $LaH_{10}$, are unconventional superconductors.

In addition, we find that both near-room-temperature superconductors, $H_3S$ and $LaH_{10}$, are subjected by strong thermodynamic fluctuations of the order parameter amplitude. We note, that this is very similar to pnictides and cuprates in which superconducting state is affected by strong fluctuations of the order parameter phase.

**Acknowledgement**

Author thanks Prof. V. E. Antonov (Institute of Solid State Physics, Russian Academy of Sciences) and Prof. E. Gregoryanz (University of Edinburgh) for valuable discussions. Author also thanks financial support provided by the state assignment of Minobrnauki of Russia (theme "Pressure" No. AAAA-A18-118020190104-3) and by Act 211 Government of the Russian Federation, contract No. 02.A03.21.0006.

# SUPPLEMENTARY INFORMAION

## for

## Classifying hydrogen-rich superconductors

E. F. Talantsev[1,2]

[1]M.N. Miheev Institute of Metal Physics, Ural Branch, Russian Academy of Sciences,

18, S. Kovalevskoy St., Ekaterinburg, 620108, Russia

[2]NANOTECH Centre, Ural Federal University, 19 Mira St., Ekaterinburg, 620002,

Russia

E-mail: evgeny.talantsev@imp.uran.ru

**Supplementary Table I.** $B_{c2}(T)$ dataset reported by Drozdov *et al.* [1] for "cooling" stage of $LaH_{10}$ subjected to pressure of $P = 150$ GPa. Values were deduced by 50% of normal state resistance criterion from Fig. 2(a) [1].

| $T$ (K) | $B_{c2}$ (T) |
|---|---|
| 241.68 | 0 |
| 241.68 | 0 |
| 241.53 | 0.2 |
| 241.26 | 0.5 |
| 239.41 | 2 |
| 236.86 | 3 |
| 236.91 | 3 |
| 233.23 | 6 |
| 229.99 | 9 |

**Supplementary Table II.** $B_{c2}(T)$ dataset reported by Drozdov *et al.* [1] for "warming" stage of $LaH_{10}$ subjected to pressure of $P = 150$ GPa. Values were deduced by 50% of normal state resistance criterion from Fig. 2(a) [1].

| $T$ (K) | $B_{c2}$ (T) |
|---|---|
| 245.1 | 0 |
| 242.5 | 3 |
| 239.0 | 6 |
| 235.7 | 9 |

**Reference**
[1] Drozdov A P, *et al* 2019 Superconductivity at 250 K in lanthanum hydride under high pressures *Nature* **569** 528-531